\documentclass[%
 aip,
 amsmath,amssymb,
 reprint,
 floatfix,%
]{revtex4-1}

\usepackage{graphicx}
\usepackage{dcolumn}
\usepackage{bm}

\usepackage[utf8]{inputenc}
\usepackage[T1]{fontenc}
\usepackage{mathptmx}

\begin{document}

\preprint{AIP/123-QED}

\title{\textit{In-situ} electromagnet with active cooling for real-time magneto-optic Kerr effect spectroscopy}

\author{A. Brozyniak}
 \altaffiliation[New affiliation:\,]{Christian Doppler Laboratory for Nanoscale Phase Transformations, Johannes Kepler University, Altenberger Str. 69, 4040 Linz, Austria}
 \affiliation{Institute of Experimental Physics, Johannes Kepler University, Altenberger Str. 69, 4040 Linz, Austria}
\author{G. Mendirek}%
\affiliation{Institute of Experimental Physics, Johannes Kepler University, Altenberger Str. 69, 4040 Linz, Austria}

\author{M. Hohage}
\affiliation{Institute of Experimental Physics, Johannes Kepler University, Altenberger Str. 69, 4040 Linz, Austria}%

\author{A. Navarro-Quezada}
\email{andrea.navarro-quezada@jku.at}
\affiliation{Institute of Semiconductor and Solid State Physics, Johannes Kepler University, Altenberger Str. 69, 4040 Linz, Austria}

\author{P. Zeppenfeld}
\affiliation{Institute of Experimental Physics, Johannes Kepler University, Altenberger Str. 69, 4040 Linz, Austria}

\date{\today}

\begin{abstract}
We present a compact \textit{in-situ} electromagnet with an active cooling system for the use in ultra-high vacuum environments. The active cooling enhances the thermal stability and increases the electric current that can be applied through the coil, promoting the generation of homogeneous magnetic fields, required for applications in real-time deposition experiments. The electromagnet has been integrated into a reflectance difference magneto-optic Kerr effect (RD-MOKE) spectroscopy system that allows the synchronous measurement of the optical anisotropy and the magneto-optic response in polar MOKE geometry. Proof of principle studies have been performed in real-time during the deposition of ultra-thin Ni films on Cu(110)-(2$\times$1)O surfaces, corroborating the extremely sharp spin reorientation transition above a critical coverage of 9 monolayers and demonstrating the potential of the applied setup for real-time and \textit{in-situ} investigations of magnetic thin films and interfaces.
\end{abstract}

\maketitle

\section{\label{Intro}Introduction}

The constantly developing field of magnetic thin films and interfaces requires reliable and sensitive methods to address and correlate structural and magnetic properties with high precision. Since the mid 1980’s the magneto-optic Kerr effect (MOKE) has been employed to investigate the magnetic properties of novel thin films and interfaces\cite{Bader:1987_JAP, Qiu:1999_RSI}. In the case of surfaces and thin films, the magnetic properties are strongly related to the structure and to the morphology and can be tuned by the choice of substrate and adsorbate materials, the film thickness or by the growth conditions. A widely investigated class of materials are inorganic/organic interfaces, where magnetic molecules are adsorbed on ferromagnetic transition metal substrates. The “spinterface” between both layers often reveals a ferro- or antiferromagnetic coupling\,\cite{Sanvito:2010_NP}. Therefore, the methods required to experimentally investigate such systems, need to simultaneously address and correlate the magnetism with the underlying morphology.

An optical technique, which elegantly incorporates the above requirements, is the combination of surface sensitive reflectance difference spectroscopy (RDS) with MOKE\,\cite{Herrmann:2006_PRB, Fronk:2009_PRB, Denk:2009_PRB}. A further improvement of this technique is the magnetic-field modulated RD-MOKE spectroscopy that allows examining morphological and magnetic changes in real-time during the deposition of magnetic molecules with high sensitivity\,\cite{Denk:2009_PRB, Mendirek:2020}. While existing spectroscopic RD-MOKE systems often employ \textit{ex-situ} generated magnetic fields\,\cite{Fronk:2009_PRB, Herrmann:2006_PRB}, it is desired to apply magnetic field \textit{in-situ} directly at the sample position and \emph{during} deposition. This requires an ultra-high-vacuum (UHV) compatible design which does not obstruct the optical path to and from the sample surface\,\cite{Qiu:1999_RSI, Bader:1987_JAP}. 

The state-of-the-art in the field of \textit{in-situ} electromagnets is mainly based on devices for very specific applications, rather than on general UHV-compatible solutions. In the context of the implementation of a transversal MOKE setup in Ref.\,\onlinecite{Froemel:2013}, the authors presented a Helmholtz-type double-coil electromagnet with a field of 5\,mT, for which a large current of 45\,A is required. A substantially improved current-to-magnetic field ratio is achieved in the yoke design electromagnet reported in Ref.\,\onlinecite{Heigl:2002_RSI}, which allows the generation of a magnetic field of 170\,mT with a current of only 4\,A. This setup, however, is not equipped with a cooling system, so its full potential can be exploited over very limited time scales or in pulse mode. An alternative solution for continuous operation is a superconducting magnet. However, superconducting electromagnets are significantly space-consuming and costly, and are usually integrated in a specific measuring system, such as a scanning tunneling microscope\,\cite{Wiebe:2004_RSI}. Until now, no universal solution for UHV chambers has been proposed.

In this work, we present a compact \textit{in-situ} UHV-compatible electromagnet with active cooling that provides a magnetic field located directly at the position of the investigated sample. A characterization of the magnetic field and of the thermal behavior of the electromagnet is provided in Section\,\ref{two}. As a proof of principle, the implementation of the device into a UHV chamber in combination with real-time modulated RD-MOKE spectroscopy is described in Section\,\ref{pop}. The results show the advantages of employing the present design for real-time investigations of the magnetic properties of growing ferromagnetic thin films and its potential to elucidate the magnetic coupling at surfaces and interfaces between organic/inorganic systems\,\cite{Gobbi:2019_JAP, Denk:2019_JAP, Cinchetti:2017_NatMat, MaMari:2015_Nature, Sanvito:2010_NP, Wende:2007_NM, Waeckerlin:2015_CC}. 

\section{Electromagnet with active cooling}
\label{two}
\subsection{\label{design}Design}
\begin{figure}
\includegraphics[width = 8 cm]{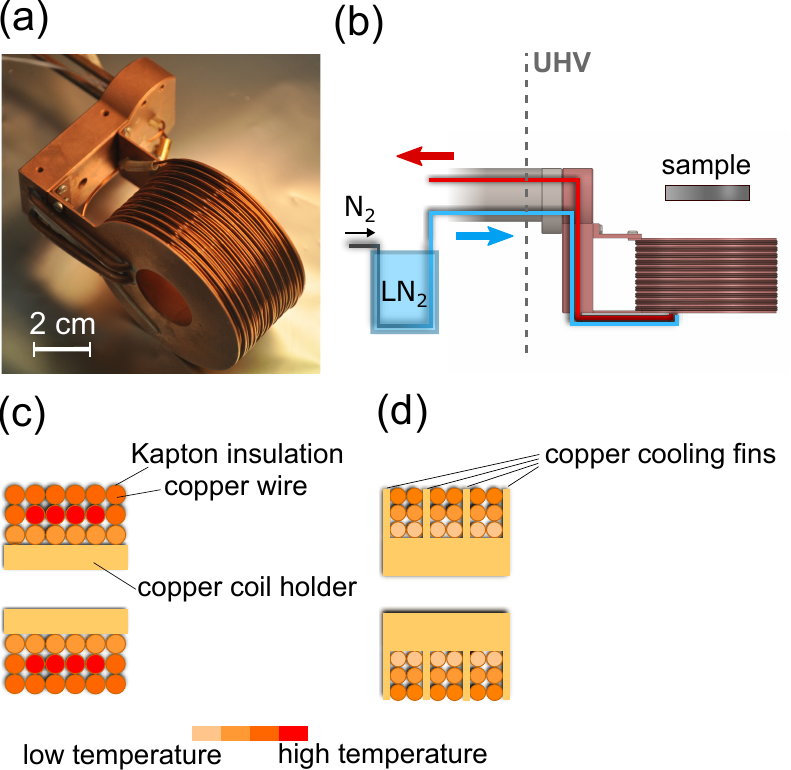}
\caption{\label{fig:magnet} (a) Image of the pancake-coil electromagnet. Schematic representations of: (b) the electromagnet and the active cooling system, (c) the wire arrangement in a conventional coil, and (d) in the present electromagnet.}
\end{figure}

An electromagnet for \textit{in-situ} applications under UHV conditions has to fulfil two main requirements. (1) In order to ensure operation in real-time during sample manipulation or thin film deposition, the electromagnet has to be positioned as close to the sample as possible, while complying with the geometrical restrictions and dimensions of the chamber. (2) The electromagnet has to be manufactured from UHV-compatible low out-gassing materials and its temperature should be kept as low as possible during operation, since otherwise the pressure inside the chamber would exceed the acceptable range for surface science experiments. For this reason, an actively cooled electromagnet was conceived, which allows applying high currents to reach large magnetic fields at limited out-gassing rates. In fact, the cooling not only reduces the thermal desorption rates, but the cooled wires also have a lower electrical resistance, which further reduces the Joule heating. 
The developed electromagnet, shown in Fig.\,\ref{fig:magnet}\,(a), features a cylindrical and regular single pancake-coil design to enhance the magnetic field homogeneity beyond that of a standard electromagnetic coil\,\cite{Brozyniak:2020_Msc}. The coil, consisting of $N\,=\,254$ windings, is wound on a copper coil holder with radially arranged cooling fins, in order to facilitate the heat flow away from the wires. The heat generation depends strongly on the electric current flowing through the coil. 

The active cooling of the electromagnet, which is the centerpiece of the development, is achieved by inlet and outlet pipes that are directly mounted into a copper heat exchanger. Cooling can be accomplished in two different ways: (i) employing an air or gaseous nitrogen (N$_2$) flow at room-temperature (RT), or using (ii) nitrogen gas pre-cooled by liquid nitrogen (LN$2$). For the second method, the inlet pipe is connected to a copper pipe array that is submerged in liquid nitrogen, as shown schematically in Fig.\,\ref{fig:magnet}\,(b). The gaseous N$_2$ is thus cooled close to its boiling temperature of 77.4\,K\,\cite{Jacobsen:1986_JPCRD}, resulting in a (partial) transition into the liquid state. 
A Kapton coated wire of 1.1\,mm in diameter was used for the coil. Kapton is a polyimide film and a widespread electrical insulator in UHV environments due to its low out-gassing rate and its thermal stability in the low and medium temperature regime\,\cite{Kittel:1994, Kapton}. An additional feature of Kapton is its low thermal conductivity in the range of 0.37-0.52\,W/(m\,$\cdot$\,K)\,\cite{Czichos:2000_c}, while commercially available copper shows a thermal conductivity coefficient of 240–380\,W/(m\,$\cdot$\,K)\cite{Bargel:2008}. Even for a Kapton film thickness of 0.05\,mm, as in the case of the used wire, this results in a major challenge concerning the heat transport from the coil. To optimize cooling efficiency, the present design consists of a pancake-like coil system, featuring thin copper cooling fins where only two wires are wound next to each other within one layer of the coil. In such an arrangement, every winding is in direct contact with one of the cooling fins, as shown in Fig.\,\ref{fig:magnet}\,(d), so that the heat can be more easily transferred to the coil holder. In contrast, the heat transfer of in a standard coil configuration, presented in Fig.\,\ref{fig:magnet}\,(c), only the wires in direct contact with the coil frame are cooled. In such an arrangement, the inner wires of the coil tend to overheat, limiting considerably the electric current through the coil and thus, the maximum achievable magnetic field. 

Another important feature of the present electromagnet is its compact design and DN\,40\,CF flange mounting, which makes it exchangeable and compatible with most UHV chambers. Due to the fact that the electromagnet is mounted on a $x$-manipulator, the positioning of the magnet inside the UHV chamber is adjustable. Furthermore, the present design allows also the incorporation of a rotational drive to the mounting, adding a further degree of freedom to the electromagnet inside the UHV chamber. 

\subsection{\label{magnetic}Magnetic characterization}
\begin{figure}
\includegraphics[width = 7.5 cm]{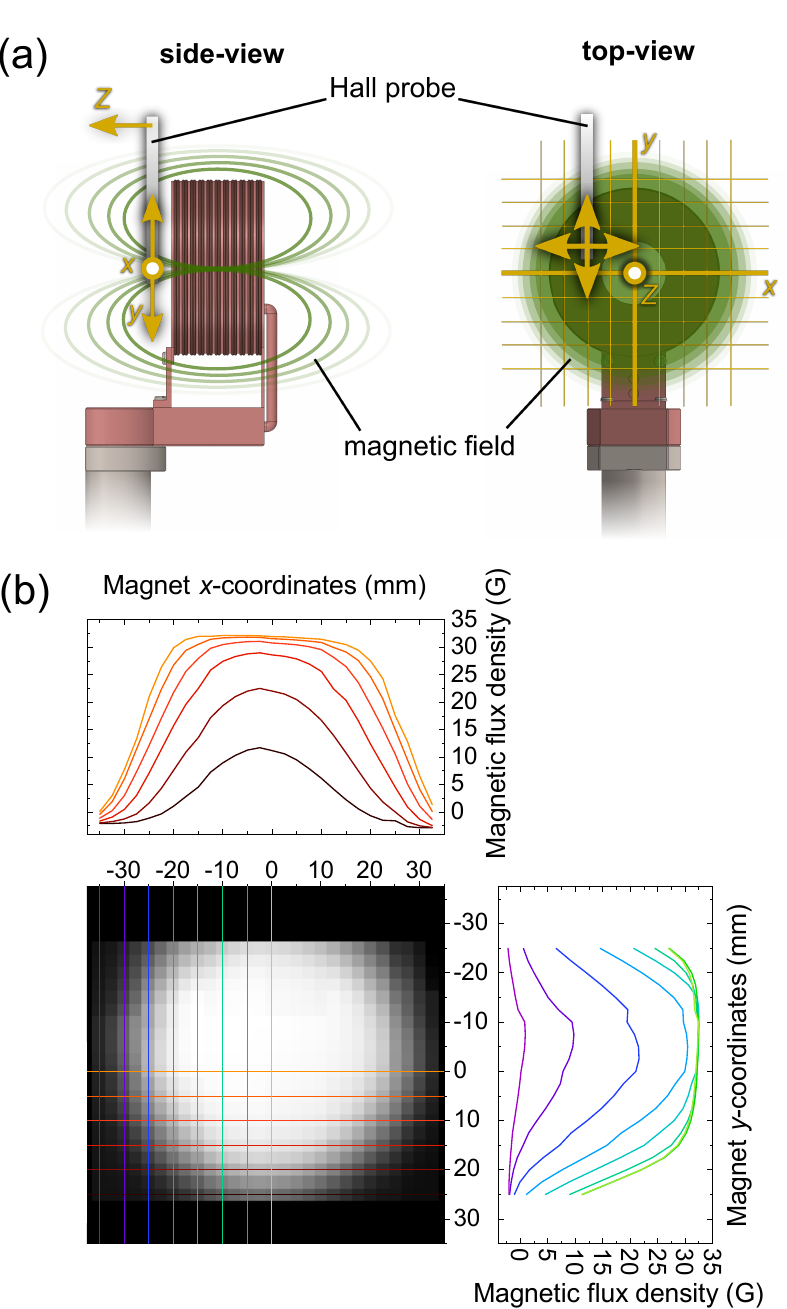}
\caption{\label{fig:magnetic}(a) Measurement configuration to determine the magnetic flux density $B(x,y,Z)$. (b) 2D-map of the magnetic flux density of the present electromagnet for a current of $I\,=\,$\,1\,A and $Z\,=$\,1.8\,mm.}
\end{figure}

The homogeneity of the magnetic field of the electromagnet was checked by measuring the magnetic flux density $B$ along the $z$-direction with a commercial Hall-probe, as shown schematically in Fig.\,\ref{fig:magnetic}\,(a). The position of the Hall-probe is controlled by optical $x$-$y$ manipulator stages. In this way, the distribution of the magnetic flux density in the $x$-$y$ plane of the magnet was determined, providing a full 2D-map of $B$, as shown in Fig.\,\ref{fig:magnetic}\,(b). Note that the distance in the $z$-direction is not measured from the center of the coil ($z$), but from the front edge of the magnet ($Z\,=\,z-L/2$), as depicted in Fig.\,\ref{fig:magnetic}\,(a). The magnetic flux density was measured every 2.5\,mm in $x$- and $y$-direction over the magnet area at $Z\,=$\,1.8\,mm and for $I\,=$\,1\,A. 

\begin{figure}
\includegraphics[width = 7.5 cm]{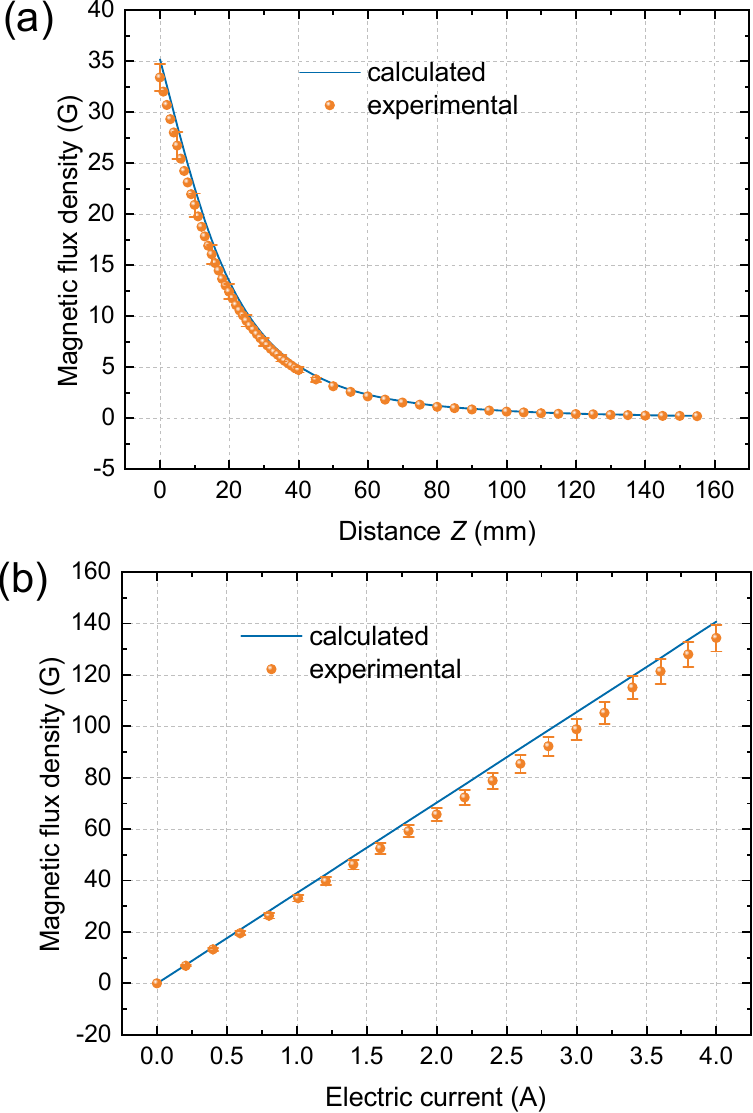}
\caption{\label{fig:magnetic2} Magnetic flux density of the electromagnet, both experimental (orange filled dots), and calculated (blue solid line): (a) as a function of the distance $Z$ from the magnet, and (b) as a function of the applied electric current.}
\end{figure}

The measured magnetic flux density along the sample axis as a function of the distance $Z$ is illustrated in Fig.\,\ref{fig:magnetic2}\,(a). The data are compared to a numerical calculation (blue solid line) based on the Biot-Savart law\,\cite{Grant:2008} considering our electromagnet as a multi-layered assembly of conducting loops: 
\begin{equation}
 B_\mathrm{axial}(Z)\,=\,B_0\cdot\,g(Z,L,r_\mathrm{o},r_\mathrm{i})
\end{equation}
with
\begin{widetext}
\begin{equation}
 g(Z,L,r_\mathrm{o},r_\mathrm{i})\,=\,\left((Z+L)\cdot\,\mathrm{ln}
 \left(
 \frac{r_\mathrm{o}\,+\,\sqrt{r_\mathrm{o}^2 + (Z+L)^2}}{r_\mathrm{i}\,+\,\sqrt{r_\mathrm{i}^2 + (Z+L)^2}}
 \right) - Z\cdot \mathrm{ln}
 \left(
 \frac{r_\mathrm{o}\,+\,\sqrt{r_\mathrm{o}^2 + (Z)^2}}{r_\mathrm{i}\,+\,\sqrt{r_\mathrm{i}^2 + (Z)^2}}
 \right)
 \cdot
 \left(
 \frac{1}{r_\mathrm{o}-r_\mathrm{i}}
 \right)
 \right)
\end{equation}
\end{widetext}
where $B_0\,=\,\mu_0\,\cdot(I\cdot L)/N$, $L$ is the length of the coil, and $r_\mathrm{o}$ and $r_\mathrm{i}$ are the outer and inner radius of the coil, respectively. For the present electromagnet, $N\,=$\,254, $L\,=$\,3.6\,cm, $r_\mathrm{o}\,=$\,3.4\,cm, $r_\mathrm{i}\,=$\,2.1\,cm and $B_0 =$\,4.9\,mT. The experimental results are in good agreement with the theoretical prediction. The magnetic flux density $B$ as a function of the applied current is depicted in Fig.\,\ref{fig:magnetic2}\,(b), showing a linear dependence in accordance with the theoretical expectations. These results have been obtained directly at the front edge of the magnet ($Z\,=$\,0). 

\subsection{\label{thermal}Thermal characterization}

In the present device, the heat transfer from the windings to the coil holder is considerably improved through the cooling fins. Moreover, the coil holder is attached to a heat exchanger with an inflow and an outflow for a cooling fluid as shown in Fig.\,\ref{fig:magnet}\,(b). As a result, the heat flow is mainly convective, rather than just relying on conductive heat transfer. The thermal properties of the magnet were measured in air conditions using a FLUKE Ti400 infrared camera. This thermal imaging camera, with 320 x 240 pixels resolution, features a temperature measurement range between -20$^{\circ}$C and +1200$^{\circ}$C and a $\pm$\,2$^{\circ}$C or $\pm$\,2\% accuracy, depending on which of the two values is larger. It offers qualitative temperature monitoring with an adjustable color scale over the whole image area, as well as an option for quantitative temperature measurement in up to 3 pixels (markers) within the image. Precise temperature acquisition with a thermal imaging camera relies on the manual setting of the emissivity (0 – 1.0). Considering the different materials on the present magnet, \textit{e.g.} bare copper for the magnet holder and Kapton for the wire insulation, and different viewing angles, the temperature measurements performed with the FLUKE Ti400 are more suited to observe a general trend in thermal behavior as well as relative temperature differences, rather than absolute and reliable temperature values. However, the results provide a good estimate of the temperature distribution within the magnet. In order to obtain a quantitative value of the temperature, an additional K-type thermocouple was attached to the electromagnet. These thermocouples are considered to be very reliable in the examined temperature range, usually featuring accuracy of $\pm$\,2.2$^{\circ}$C or $\pm$\,0.75\%\,\cite{ThermoK}.
\begin{figure}
\includegraphics[width = 7.5 cm]{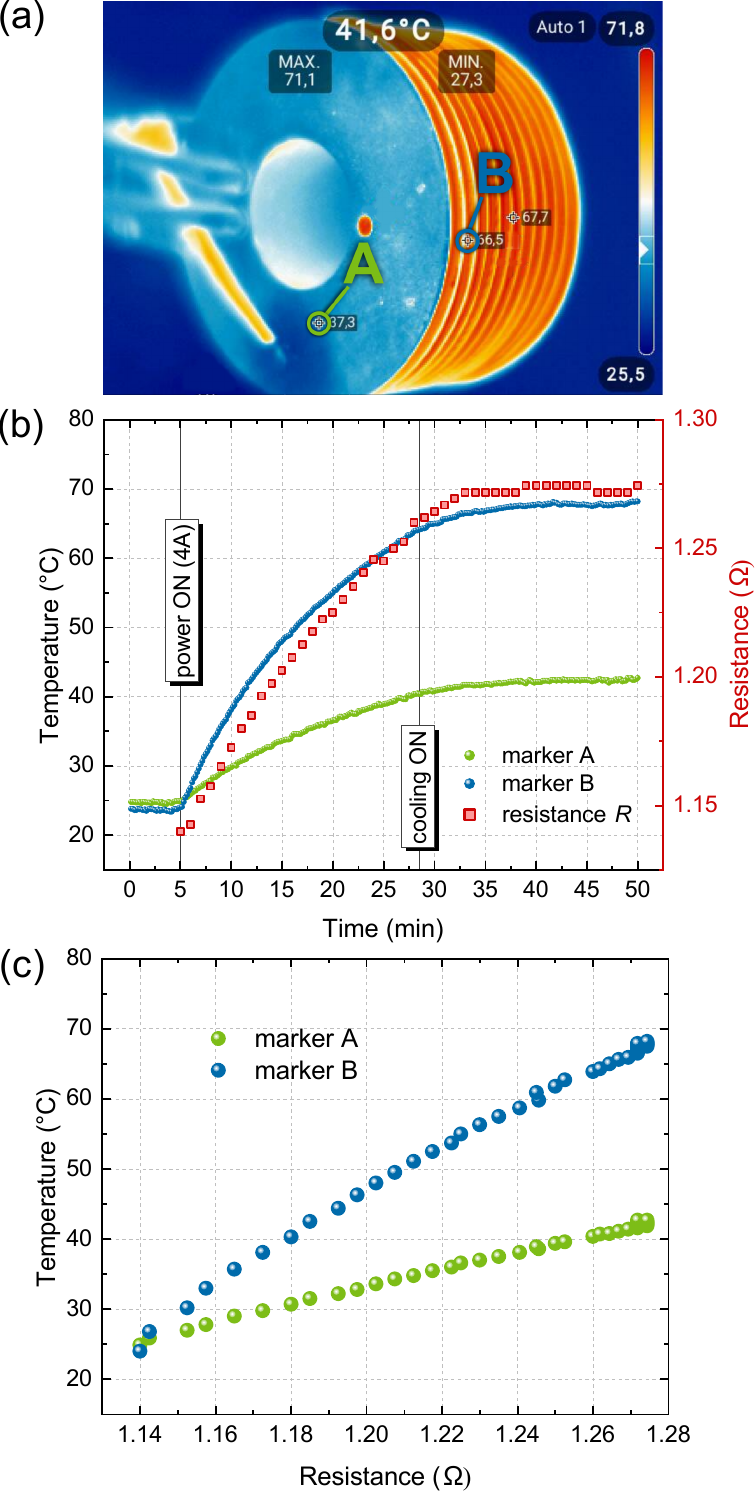}
\caption{\label{fig:thermal} (a) Infrared image of the electromagnet with the indicated positions of the markers A and B. (b) Evolution of the temperature and the resistance as a function of time upon cooling with gaseous N$_2$ at RT. (c) Temperature as a function of resistance for markers A and B.}
\end{figure}

\begin{figure}
\includegraphics[width = 7.5 cm]{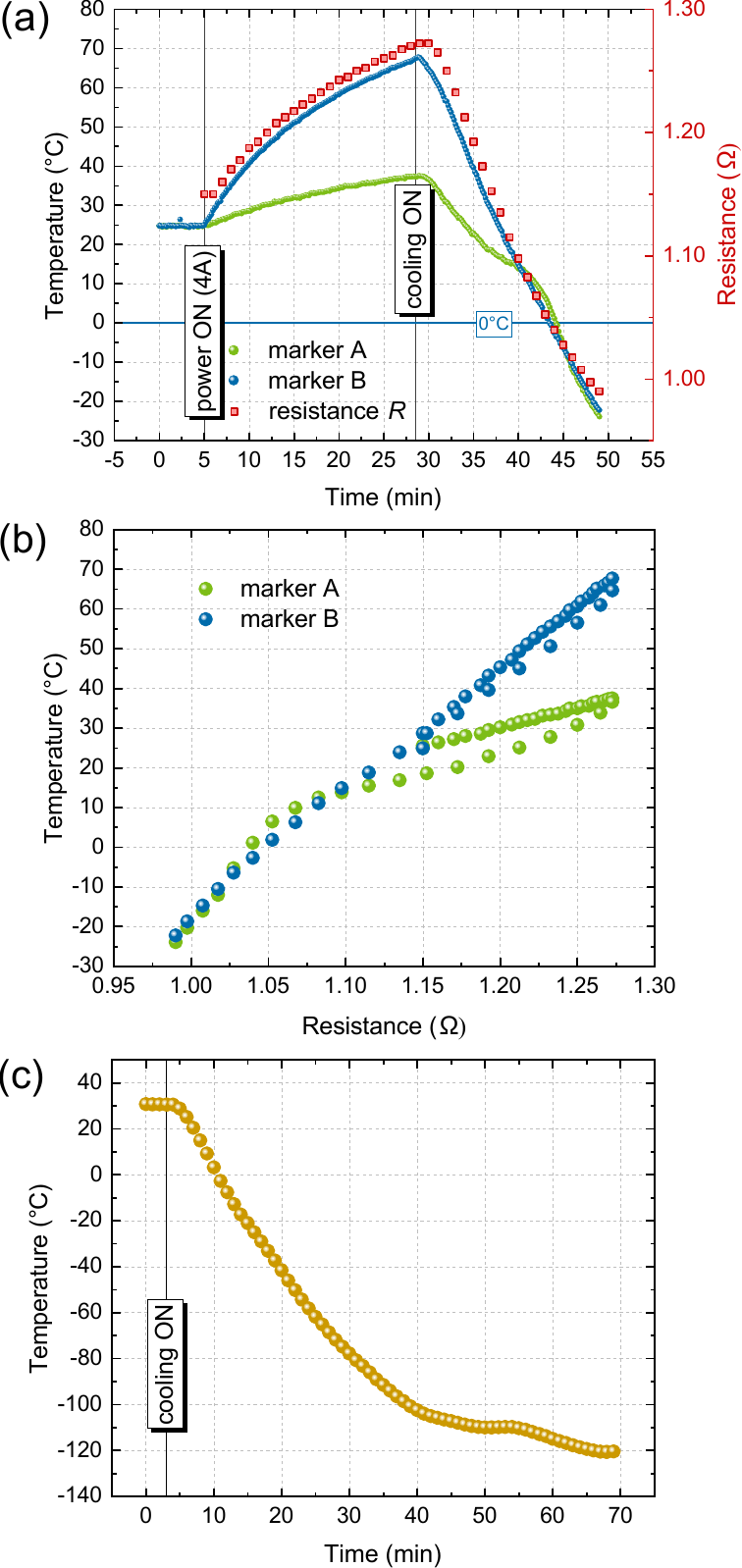}
\caption{\label{fig:ln2}(a) Evolution of the temperature and the resistance as a function of time while cooling with LN$2$. (b) Temperature as a function of resistance, (c) Temperature evolution of the electromagnet upon cooling with LN$2$ in UHV.}
\end{figure}

The infrared camera was used to monitor the heating process, by taking an image every 10\,s, while the electromagnet was operated with a current of $I\,=$\,4\,A. As soon as the temperature of the electromagnet reached approximately 70$^{\circ}$C, the cooling was turned on. The temperature of the magnet was measured quantitatively using two available markers (A and B) of the infrared camera, as displayed in Fig.\,\ref{fig:thermal}\,(a).  
Additionally, the temperature of the magnet was also estimated by measuring the change in electrical resistance during the heating process. The resistivity $\rho$ of the coil rises with increasing temperature, according to the Taylor-expansion
\begin{equation}
 \rho\,=\,\rho_\mathrm{20}\cdot(1+\alpha_\mathrm{20}(T-20^\circ\,\mathrm{C}) + \beta_\mathrm{20}(T-20^\circ\,\mathrm{C})^2 + ...)
\end{equation}
where $\rho_{\mathrm{20}}\,=$\,0.01786\,$\Omega\,\cdot$mm$^2$/m is the resistivity of copper at $T\,=$\,20$^{\circ}$C and $\alpha_{\mathrm{20}}$ and $\beta_{\mathrm{20}}$ are the temperature coefficients of the copper wire at 20$^{\circ}$C\,\cite{Stiny:2015}. In the measured temperature range below $T\,=$\,150$^{\circ}$C it is sufficient to use only the first-order linear temperature coefficient $\alpha_{\mathrm{20}}\,=$\,3.93$\cdot$10$^{-3}$\,K$^{-1}$. Considering the relation between resistance $R$ and resistivity $\rho$ being $R\,=\,\rho\,l/A$, where $l$ is the length and $A$ is the cross-section area of a wire, the above formula transforms into
\begin{equation}
 R\,=\,R_\mathrm{20}\cdot(1+\alpha_\mathrm{20}(T-20^\circ\,\mathrm{C}).
\end{equation}
In the following experiments the values for the resistance were obtained by measuring the applied voltage $U$ at a constant current $I\,=$\,4\,A und using Ohm’s law, $R\,=\,U/I$. 
The evolution of the temperature and the resistance upon cooling with nitrogen (N$_2$) gas at RT are depicted in Fig.\,\ref{fig:thermal}\,(b). It can be noticed that the coil holder reacts upon cooling by keeping its temperature in a rather constant range, making the setup partly suitable for RT cooling fluids. As can be seen in Fig.\,\ref{fig:thermal}\,(c), the temperature scales linearly with the resistance, which is an indication for a homogeneous temperature distribution. 

The behavior of the temperature and the resistance upon cooling with LN$2$ is shown in Fig.\,\ref{fig:ln2}\,(a). It can be clearly stated that the coil is cooling down very homogeneously and the time needed to reach 0$^{\circ}$C is 15\,min. Except for marker A, which is not placed on the coil, all the measured temperatures scale linearly with the resistance over the whole range in Fig.\,\ref{fig:ln2}\,(b), providing further evidence for a highly homogeneous heating and cooling of the pancake-coil.
The electromagnet was further tested under vacuum conditions while cooling with LN$2$. No current was applied during the experiments. The temperature was measured solely with the thermocouple on the coil holder, since infrared cameras are not suitable for measurements through UHV chamber windows. The results are illustrated in Fig.\,\ref{fig:ln2}\,(c), confirming that the employed cooling method is suitable for UHV. 

\section{\label{pop}Proof of principle}

\begin{figure}
\includegraphics[width = 8 cm]{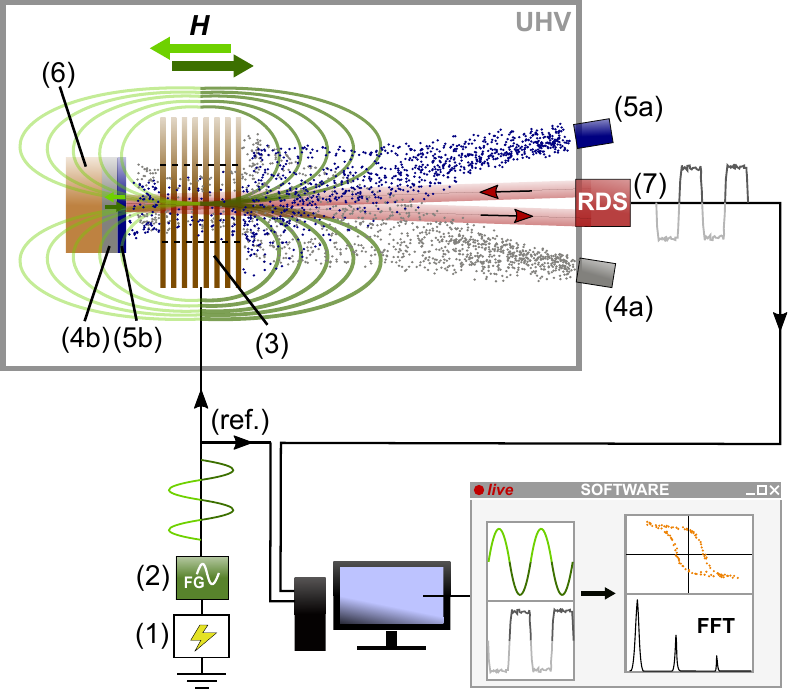}
\caption{\label{fig:setup} Schematic representation of the employed RD-MOKE setup with modulated external magnetic field provided by the newly developed \textit{in-situ} UHV electromagnet with active cooling.}
\end{figure}

The proof of principle of the \textit{in-situ} electromagnet with active cooling has been carried out in a UHV chamber equipped with a RD-MOKE spectroscopy setup, as illustrated in Fig.\,\ref{fig:setup}. The electric current from the power source (1) is modulated by a function generator (2) into a sinusoidal waveform. The modulated current, which is also redirected as a reference signal (ref.) to the computer (PC), generates a modulated magnetic field $H$ (light and dark green arrows) in the pancake-coil electromagnet (3). During the experiments, the electromagnet was cooled by LN$2$. A current of 1\,A corresponds to a magnetic field of $\mu_0H\approx$\,3\,mT at the sample position. During operation of the electromagnet with current amplitudes of up to 6\,A (corresponding to $\mu_0H=$\,18\,mT at the sample surface), the base pressure in the vacuum chamber was less than 2$\times$10$^{-10}$\,mbar and the temperature of the electromagnet during operation and cooling with LN2 stabilized at -20$^\circ$C. Operation at constant currents of up to 7\,A ($\mu_0H=$\,21\,mT) is also possible, preferably for shorter periods of time, \textit{e.g.} in pulse mode or for recording of single hysteresis curves.

In the chamber either metals (4b) or organic materials (5b) can be deposited from the respective deposition sources (4a) and (5a) onto the substrate (6), in this case consisting of a Cu(110) single crystal. In the geometry employed here, when the magnetic easy axis of the film switches from in-plane to out-of-plane as a result of the spin reorientation transition (SRT), the magnetic moment follows the modulated field of the magnet, yielding a sizeable signal in the polar RD-MOKE set-up\,\cite{Herrmann:2006_PRB, Denk:2009_PRB}. The phase modulated information from the light reflected from the sample surface at a small angle of about 2$^\circ$ measured with the RDS (7), is transformed into an intensity modulated signal with an analyzer\,\cite{Weightman:2005_RPP}. The resulting modulated RD signal and the reference signal (ref.) are recorded and processed in the PC. The front end displays the real-time view of the magnetization curve and a Fourier decomposition of the modulated RD signal, from which other measurement related parameters can be derived\,\cite{Mendirek:2020}. 

The oxygen pre-coverage on the Cu(110) surface is required to overcome the rather disfavored 3D-type growth mode of Ni on the pristine surface\,\cite{Herrmann:2006_PRB, Wahl:2003_pssc}. The pre-adsorbed oxygen acts as a surfactant and induces a layer-by-layer growth mode over an extended coverage regime\,\cite{Denk:2009_PRB}. The oxygen atoms float on top of the grown Ni layers and induce a (2$\times$1)\,O reconstruction at the surface of the pseudomorphic Ni(110) film. The shape anisotropy initially favors an in-plane orientation of the magnetic easy axis, which switches to an out-of-plane orientation at a Ni film thickness of 9 monolayers (ML)\,\cite{Denk:2009_PRB, Herrmann:2006_PRB}. This sharp SRT arises from the increasing magneto-elastic anisotropy as a consequence of the accumulating strain during the pseudomorphic growth of the Ni film.
\begin{figure}
\includegraphics[width = 7.8 cm]{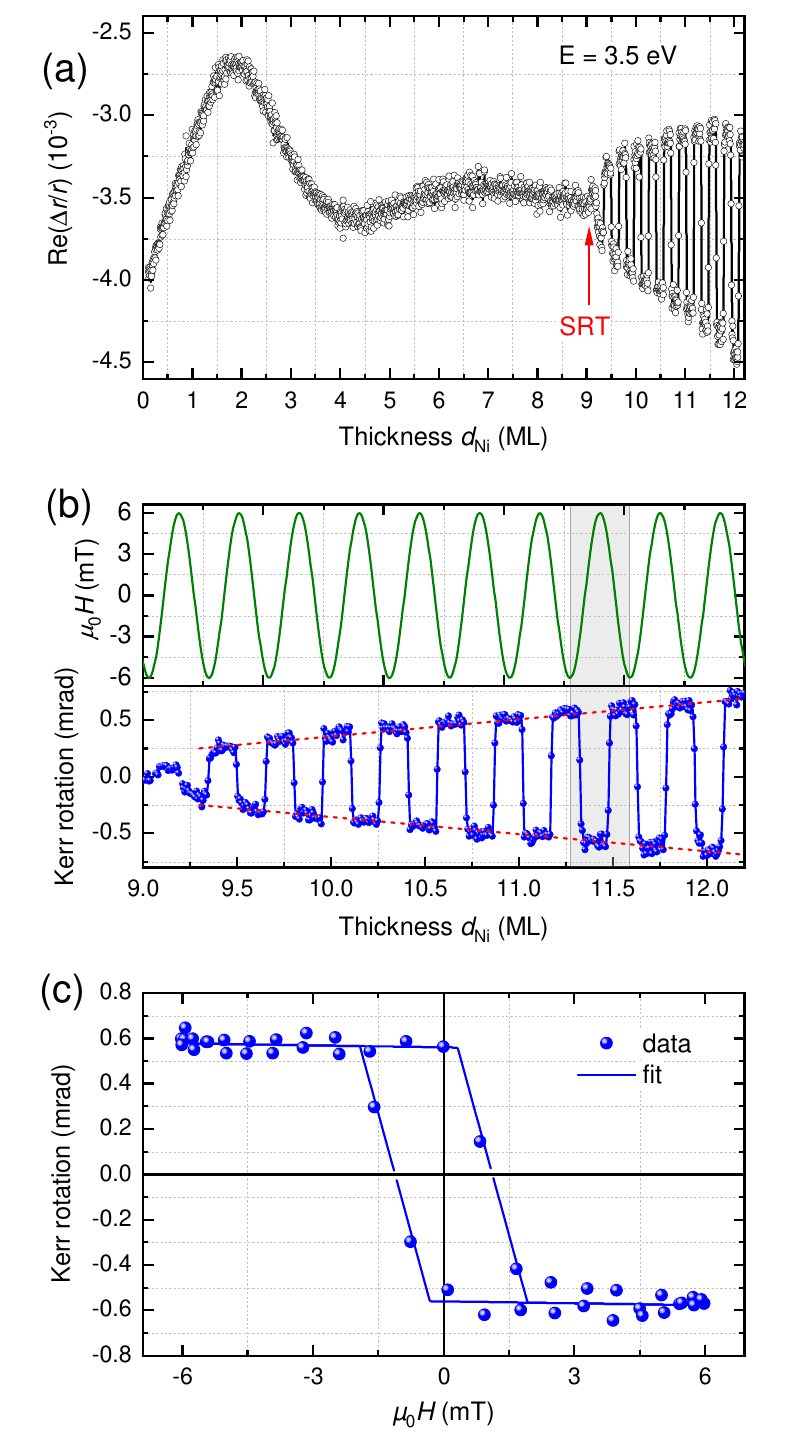}
\caption{\label{fig:rdmoke} (a) RD transient at a photon energy of $E\,=$\,3.5\,eV recorded as a function of the Ni film thickness $d_\mathrm{Ni}$ during the deposition of Ni while applying a modulated magnetic field of $\pm$\,6\,mT. The SRT at a coverage of 9\,ML is marked with a red arrow. (b) Comparison of the applied external field $\mu_0H$ and the Kerr rotation as a function of $d_\mathrm{Ni}$ after the SRT, showing the linear dependence of the Kerr rotation with increasing layer thickness. (c) Hysteresis curve extracted from the shaded area in (b): data (symbols) and fit (solid line).}
\end{figure}

The sudden onset of the out-of-plane magnetization is reproduced in Fig.\,\ref{fig:rdmoke}\,(a), where the RD signal at a photon energy of $E\,=$\,3.5\,eV is monitored during the deposition of 12.3\,ML of Ni on a Cu(110)-(2$\times$1)O surface. The Ni was deposited with a rate of 0.37\,ML/min and with the sample held at RT. A sinusoidally modulated magnetic field with an amplitude $\mu_0H=$\,6\,mT and a frequency $f=$\,0.02\,Hz, as shown in the upper panel of Fig.\,\ref{fig:rdmoke}\,(b), was generated by the \textit{in-situ} pancake-coil electromagnet applied normal to the surface during deposition. The onset of the sharp SRT at 9\,ML is clearly identified, since at this point, the RD signal starts oscillating in response to the applied magnetic field. The oscillating RD-MOKE signal after the SRT, presented in detail in the lower panel of Fig.\,\ref{fig:rdmoke}\,(b), quickly takes the shape of a square-wave signal, which is an indication for out-of-plane ferromagnetism\,\cite{Denk:2019_PhD, Aspelmeier:1995_JMMM}. From this oscillating RD-MOKE signal, magnetization curves, more precisely, the Kerr rotation, which is proportional to the sample magnetization, as a function of the external magnetic field, can be extracted. As an example, the hysteresis curve shown in Fig.\,\ref{fig:rdmoke}\,(c), was obtained from the shadowed area in Fig.\,\ref{fig:rdmoke}\,(b), corresponding to a full cycle of the external magnetic field. 

In this way, it is possible to distinguish between a paramagnetic or ferromagnetic response and to follow its evolution during growth. The magnetization curves can further be fitted to model functions, as shown by the solid blue line in Fig.\,\ref{fig:rdmoke}\,(c). From these fits, important parameters such as the Kerr amplitude, the remanent Kerr rotation and the coercive field can be determined in real-time as a function of coverage and sample temperature\,\cite{Mendirek:2020}.

\section{Conclusions}
 A compact \textit{in-situ} pancake-coil electromagnet for operation in UHV environments has been developed with an active cooling system to reduce the challenges related to the Joule heating and the out-gassing occurring in UHV. The functionality of the present electromagnet was successfully characterized and tested both \textit{ex-situ} and \textit{in-situ}. The advantages of the present electromagnet clearly stand out from comparable solutions for UHV environments. Due to its \textit{in-situ} nature, the magnet requires much lower currents for achieving the same magnetic field at the sample position than \textit{ex-situ} setups. Moreover, because of the proximity to the sample, the field is quite homogeneous and not distorted by any additional chamber components. As a result of the active cooling of the electromagnet, high currents of 7\,A can be applied, yielding magnetic field strengths as high as 21\,mT in continuous mode. Besides, its energy- and space-saving characteristics, the electromagnet is mounted on a DN\,40\,CF flange and is compatible to any standard UHV chamber, which makes it a simple and powerful add-on to UHV systems where a magnetic field is required. 
 
 The here developed electromagnet was combined with an existing RD-MOKE setup and the first \textit{in-situ} modulated RD-MOKE experiments have been realized. The modulation of the magnetic field allows monitoring the magnetic properties in real-time as demonstrated here for the case of Ni thin films deposited on a Cu(110)-(2$\times$1)O surface. The ease and beauty of the field modulation technique, can be fully exploited using a suitable \textit{in-situ} electromagnet, which thanks to its design, delivers highly homogeneous magnetic fields up to 21\,mT. Such magnetic field strengths would also be sufficient for real-time investigations of hybrid organic/inorganic magnetic systems\,\cite{MaMari:2015_Nature, Denk:2019_JAP, Wende:2007_NM, Waeckerlin:2015_CC}.\\

The following article has been submitted to Review of Scientific Instruments. After it is published, it will be found at https://aip.scitation.org/journal/rsi. 

\begin{acknowledgments}
The authors acknowledge the financial support by the Austrian Science Fund (FWF) Project no.\,V-478-N36. Technical support by Robert Leimlehner, Ewald Fink and Konrad Fragner is greatly acknowledged. The authors would also like to thank Margherita Matzer for her support with the Austrian Patent Office. 
\end{acknowledgments}

%

\end{document}